\documentclass[a4paper,twocolumn,aps,prd,nolongbibliography,superscriptaddress,showpacs,showkeys,amsmath,amssymb,floatfix,nofootinbib]{revtex4-1}
\usepackage{lmodern}
\usepackage{yfonts}
\usepackage{graphicx}

\usepackage[T1]{fontenc}
\usepackage[utf8]{inputenc}
\usepackage[usenames]{xcolor}
\usepackage[unicode=true,pdfusetitle,
 bookmarks=true,bookmarksnumbered=true,bookmarksopen=true,bookmarksopenlevel=1,
 breaklinks=false,pdfborder={0 0 0},backref=false,colorlinks=true]
 {hyperref}
\hypersetup{
 citecolor=blue,filecolor=blue,linkcolor=blue,urlcolor=blue}
\makeatletter

\pdfpageheight\paperheight
\pdfpagewidth\paperwidth

 \@ifundefined{textcolor}{}
 {%
   \definecolor{BLACK}{gray}{0}
   \definecolor{WHITE}{gray}{1}
   \definecolor{RED}{rgb}{1,0,0}
   \definecolor{green}{rgb}{0,0.7,0}
   \definecolor{BLUE}{rgb}{0,0,1}
   \definecolor{CYAN}{cmyk}{1,0,0,0}
   \definecolor{MAGENTA}{cmyk}{0,1,0,0}
   \definecolor{YELLOW}{cmyk}{0,0,1,0}
      \definecolor{myred}{rgb}{0.7,0,0.1}
 }

\renewcommand{\[}{\begin{equation}}
\renewcommand{\]}{\end{equation}} 
\usepackage{array}
\makeatother

\newcommand{\corr}[1]{\left\langle #1 \right\rangle}
\newcommand{\Hcal}{\mathcal{H}}
\newcommand{\ds}{\mathrm{d}}
\newcommand{\gs}{\mathrm{g}}
\newcommand{\ms}{\mathrm{m}}
\newcommand{\rb}{\boldsymbol{r}}
\newcommand{\nb}{\boldsymbol{n}}
\newcommand{\kb}{\boldsymbol{k}}
\newcommand{\wb}{\boldsymbol{w}}
\newcommand{\xb}{\boldsymbol{x}}
\newcommand{\omb}{\boldsymbol{\omega}}
\newcommand{\kat}{\hat{k}}

\begin{document}

\title{Redshift space distortions from vector perturbations}

\author{Camille Bonvin}

\affiliation{D\'epartment de Physique Th\'eorique and Center for Astroparticle Physics,
Universit\'e de Gen\`eve, Quai E. Ansermet 24, CH-1211 Gen\`eve 4, Switzerland}

\author{Ruth Durrer}

\affiliation{D\'epartment de Physique Th\'eorique and Center for Astroparticle Physics,
Universit\'e de Gen\`eve, Quai E. Ansermet 24, CH-1211 Gen\`eve 4, Switzerland}

\author{Nima Khosravi}

\affiliation{Department of Physics, Shahid Beheshti University, G.C., Evin, Tehran 19839, Iran}

\author{Martin Kunz}

\affiliation{D\'epartment de Physique Th\'eorique and Center for Astroparticle Physics,
Universit\'e de Gen\`eve, Quai E. Ansermet 24, CH-1211 Gen\`eve 4, Switzerland}

\author{Ignacy Sawicki}

\affiliation{CEICO, Fyzikální ustáv Akademie věd ČR, Na Slovance 2, 182 21 Praha 8, Czechia}

\begin{abstract}
We compute a general expression for the contribution of vector perturbations to the redshift space distortion of galaxy surveys. We show that they contribute to the same multipoles of the correlation function as scalar perturbations and should thus in principle be taken into account in data analysis. We derive constraints for next-generation surveys on the amplitude of two sources of vector perturbations, namely non-linear clustering and topological defects. While topological defects leave a very small imprint on redshift space distortions,  we show that the multipoles of the correlation function are sensitive to vorticity induced by non-linear clustering. Therefore future redshift surveys such as DESI or the SKA should be capable of measuring such vector modes, especially with the hexadecapole which appears to be the most sensitive to the presence of vorticity.
\end{abstract}
\maketitle

\section{Introduction}

The local properties of space-time in general relativity are described by the metric tensor $g_{\mu\nu}$. In cosmology, the metric tensor is expected to be close to the Friedmann-Lema\^itre-Robertson-Walker (FLRW) form that describes a perfectly homogeneous and isotropic universe, with small fluctuations. A key goal for observational cosmology is to constrain the properties of these fluctuations.

Small fluctuations can be decomposed into scalar, vector and tensor degrees of freedom and, at the level of linear perturbation theory these degrees of freedom do not mix. Scalar metric perturbations are linked to density perturbations and gradient velocity fields and describe gravitational clustering. Vector perturbations are related to the vorticity of the velocity field and frame dragging. And tensor perturbations describe gravitational waves and tensor anisotropies of spacetime. The formation of structure in the Universe thus predominantly creates scalar metric perturbations, which is why observational cosmology has focused primarily on these. Additionally, the dominant paradigm for generating the initial perturbations in the Universe, inflation, is generically based on a scalar field and creates only scalar and tensor perturbations. Moreover both vector and tensor perturbations redshift away if they are not sourced continuously by anisotropic stresses which, in concordance cosmology, are very small in the late Universe.

However, once structure formation becomes non-linear, the separation into scalars, vectors and tensors breaks down, so that at late times and on small scales vector perturbations are necessarily generated. Even though vorticity is conserved in a perfect fluid, dark matter is free-streaming and gravitational clustering leads to shell crossing which induces velocity dispersion and therefore vorticity~\cite{Aviles:2015osc,Piattella:2015nda,Cusin:2016zvu}. Observations of coherent angular velocity out to scales of up to 20 Mpc at redshift $z=1$ have lately been reported~\cite{Taylor:2016rsd}. It is not clear whether such large coherence scales can be reached for vector perturbations generated by shell crossing within standard $\Lambda$CDM.

Other mechanisms such as topological defects~\cite{Durrer:2001cg,Daverio:2015nva,Lizarraga:2016onn}, magnetic fields~\cite{Durrer:1998ya}, inflation with vector fields~\cite{Ford:1989me,Golovnev:2008cf} or vector-field-based models of modified gravity~\cite{Jacobson:2000xp,Zlosnik:2006zu,Heisenberg:2014rta,Tasinato:2014eka} can generate vector perturbations throughout the history of the Universe and on a wide range of scales.  If they persist until late times, or are even generated there, the presence of such vector perturbations could `pollute' the measurement of the scalar degrees of freedom and spoil precision cosmology with future large surveys if they are not properly taken into account. On the other hand if they are measured and characterized, they can turn into a signal instead of a source of systematic uncertainty, and improve our understanding of the Universe.

A large part of the effort of measuring vector-type deviations of the metric has been focused on the Cosmic Microwave Background (CMB). The approaches can be grouped into three categories: (i) introducing dynamical vector degrees of freedom in the early universe, but maintaining isotropy and homogeneity of the background. One can then either maintain statistical isotropy and homogeneity of the perturbations or allow for statistically anisotropic perturbations. In that case, there is a new contribution to scalar and tensor fluctuations, but symmetry does not allow for a generation of a strong vector signal \cite{Lim:2004js}. Nonetheless an effect could in principle be measured through B-modes in the CMB polarisation \cite{Nakashima:2011fu}. Alternatively (ii), one can deform the isotropy of the cosmological background and therefore constrain its anisotropy, while keeping the matter content standard, ensuring that this anisotropy decays with time \cite{Saadeh:2016bmp}. Finally (iii), one can posit a mechanism to introduce an anisotropy directly in the primordial power spectrum (through some interactions in the early universe, e.g.\ \cite{Ackerman:2007nb}). One then tries to look for `anomalies' in the CMB, such as in, for example,~\cite{Ade:2015hxq}. One can also look for this primordial signal in galaxy surveys, \cite{2010JCAP...05..027P,Jeong:2012df,Shiraishi:2016wec,Sugiyama:2017ggb}.

In this paper we will focus on the impact of the vector modes in the peculiar velocity field of galaxies, irrespective of their origin and thus on the redshift-space distortions (RSD) observed in galaxy surveys. In Section~\ref{sec:vec_gen} we describe the vector contribution, and how we model it. Section \ref{sec:rsd} computes the impact of vector perturbations on redshift-space distortions, providing the general expression for vector RSD and showing that they  also contribute to the monopole, quadrupole and hexadecapole of the galaxy correlation function, adding to the effect of the scalar fluctuations. We compute estimates of the detectability of these contributions with galaxy surveys in Section \ref{sec:fisher}, before presenting our conclusions.

\section{Vector Contribution to Galaxy Velocities}

\label{sec:vec_gen}

\subsection{Effect of Vector Contribution\label{sec:basics}}

We assume that our Universe is described by a perturbed FLRW metric, which we gauge-fix so that
\begin{align}
\mathrm{d}s^2 = a^2(\tau)&\Big[-(1+2\Psi)\mathrm{d}\tau^2 - \Sigma_i \mathrm{d}\tau\mathrm{d}x^i + \label{eq:metric}\\
&\,+(1-2\Phi)\delta_{ij}\mathrm{d}x^i\mathrm{d}x^j\Big] \,. \notag
\end{align}
Here $\Phi$ and $\Psi$ are the standard Newtonian scalar potentials, and $\Sigma_i$ is a pure vector fluctuation  $\partial_i\Sigma^i=0$, related to frame dragging.%
\footnote{We have fixed the gauge such that the $0i$ component of the metric has no scalar perturbations and so that the vector part of the $ij$ component vanishes. We also neglect gravitational waves (tensor perturbations).} %
We define $\mathcal{H}\equiv a'/a$ to be the conformal Hubble parameter.

The vector field $\Sigma_i$ is characterised by an amplitude and a direction. If the vector is purely time-dependent, it can be reabsorbed by a change of coordinates. We will assume that the vector field is described by a fluctuating amplitude taken from a homogeneous and isotropic distribution with the statistics of the direction described by a  tensor $W_{ij}$, defined below.

We are interested in the imprint of the vector field on the two-point correlation function of galaxies in redshift-space. The general velocity field for galaxies located at position $\boldsymbol{r}$ at conformal time $\tau$, $v^i(\boldsymbol{r},\tau)$, can be decomposed into a scalar (potential) part, $v$, and a pure vector part, $\Omega^i$, with $\partial_i\Omega^i =0$, 
\begin{equation}
v^{i}\equiv \partial^{i}v+\Omega^{i} \,.\label{eq:vel}
\end{equation}
The gauge-invariant relativistic vorticity~\cite{Rbook} is actually $a(\Omega_i-\Sigma_i)$.  This quantity is obtained by lowering the index from $\Omega^i$ with the perturbed metric. The relativistic vorticity is often denoted $\Omega_i$ (e.g.\ in~\cite{Lu:2008ju,Rbook}) and it is an additional rotational velocity over and above the frame-dragging effect. Here, however we will focus on $\Omega^i$ as it is the velocity with an upper index that is relevant for us. Note that this difference is only relevant on large scales. Well inside the horizon, the contribution from $\Sigma_i$ in concordance cosmology can by neglected whenever $\Omega_i$ does not vanish.

The galaxies are typically assumed to move on timelike geodesics of the metric, i.e.\ to obey Euler's equation.
To first order in perturbation theory the Euler equation for perfect fluids implies
\begin{equation}
\dot{\Omega}_i - \dot{\Sigma}_i  + \mathcal{H} (\Omega_i-\Sigma_i) = 0 \,. \label{eq:vecgeo}
\end{equation}
This means that geodetic motion will cause the vorticity to redshift away with only the frame-dragging  effect remaining, if it is sourced through gravity. Indeed the vector part of the first-order Einstein equations is given by
\begin{align}
\Delta \Sigma_i &= 16\pi G_\text{N}a^2 \delta T^{\,0}_{(V)i}\, , \\
\dot{\Sigma}_{(i,j)}&+2\mathcal{H} \Sigma_{(i,j)}=-8\pi G_\text{N}a^2 \delta T^{\,i}_{(V)j}\, ,
\end{align}
where $\delta T^{\,\alpha}_{(V)\beta}$ is the vector part of the perturbation of the energy-momentum tensor, which depends generally on the velocity, the anisotropic stress and the metric perturbation itself.

The relativistic vorticity is conserved for a perfect fluid also within General Relativity at all orders~\cite{Lu:2008ju}.
In a perfect fluid therefore, if vector perturbations vanish initially, there is no vorticity generation and $\Omega_i-\Sigma_i=0$ at all times. 

In the perfect-fluid approximation of concordance cosmology, there are no vector degrees of freedom, or sources of anisotropic stress, which would have a significant effect on the gravitational field and therefore on peculiar velocities. In the real Universe, however, dark matter (or galaxies) are not truly a perfect fluid. They are free-streaming, i.e.\ they move on geodesics, and as soon as shell crossing occurs, velocity dispersion can no longer be neglected and vorticity is generated for the fluid of the averaged dark matter particles (or galaxies).

In this paper, we are interested in understanding how galaxy correlations can be used independently of other probes to constrain the existence of any vector fluctuations at late times. We will discuss in detail what the current expectations for vorticity within the $\Lambda$CDM paradigm are and to what extent it is possible to measure it.

\subsection{Statistical Properties of Vector Fluctuations}

In order to compute the two-point correlation function of galaxies (2PCF), we need a model for the two-point correlation of the vector velocity, $\corr{\Omega_i\Omega_j}$ and its cross-correlation with the dark matter overdensity $\corr{\delta_\ms\Omega_i}$. We will characterise their structure in Fourier space, with our Fourier transform convention defined by
\[
f(\kb) = \int d^3 x f(\xb) e^{-i \kb . \xb} \, .
\]
We assume that the power spectrum of the amplitude of the fluctuations obeys statistical isotropy and homogeneity, i.e.\ that it depends only on the magnitude of the wave number $k\equiv |\mathbf{k}|$.

\begin{enumerate}
\item The auto-correlation of the vector field takes the form
\begin{align}
\langle\Omega_{i}(\kb)\Omega_{j}(\kb')\big\rangle = (2\pi)^{3}\delta^{(3)}(\kb+\kb')\times \nonumber\\
\left[W_{ij}P_{\Omega}(k) +i\alpha_{ij}P_A(k)\right] \,,
\end{align}
where $P_{\Omega}(k)$  and $P_A(k)$ contain information about the amplitude of the vector field, and $W_{ij}$ and $\alpha_{ij}$ are, respectively, symmetric and anti-symmetric tensors, that encode the dependence on direction.
Since $\Omega_i$ is a pure vector field, $W_{ij}$ and $\alpha_{ij}$ must satisfy $k^i W_{ij}=k^j W_{ij}=k^i\alpha_{ij}=k^j \alpha_{ij}=0$. 
The $P_A$-term is parity odd while the $P_\Omega$-term is parity even. If no parity violating processes occur in the Universe we may set $P_A=0$.
The most general form for $W_{ij}$ is 
\begin{align}
\hspace{0.7cm}W_{ij}=~&\frac{\omega}{2}\left(\delta_{ij}-\hat{k}_{i}\hat{k}_{j}\right)+ \label{eq:wij}\\
&+\bar{\omega}_{ij}-\bar{\omega}_{il}\hat{k}^{l}\hat{k}_{j}-\bar{\omega}_{lj}\hat{k}^{l}\hat{k}_{i}+
\bar{\omega}_{lm}\hat{k}^{l}\hat{k}^{m}\hat{k}_{i}\hat{k}_{j} \,,\nonumber
\end{align}
with an \emph{arbitrary} constant symmetric tensor, which we have already decomposed into its trace $\omega$ and trace-free part $\bar{\omega}^i_i=0$. 
As usual $\hat{\kb}$ denotes the unit vector in the direction of the vector $\kb$.
The  tensorial form for $\alpha_{ij}$ is completely fixed by anti-symmetry and transversality, 
\begin{equation}\label{eq:aij}
\hspace{0.3cm}\alpha_{ij}= \alpha\epsilon_{ijm}\hat k^m \,.
\end{equation}

The first line of~\eqref{eq:wij} respects statistical homogeneity and isotropy, whereas the second line is non-zero only when isotropy is violated. In what follows, we  absorb the trace $\omega$ into the normalisation of the power spectrum $P_\Omega$. The only possible parity odd term given in  \eqref{eq:aij} is statistically isotropic.

In general, since $\bar\omega_{ij}$ is symmetric, it can be decomposed into a sum of three tensor products of its orthonormal eigenvectors $\bar\omega_i^{I}$, 
\[
\bar\omega_{ij}=\sum_{I=1}^{3}\lambda_{I}\bar\omega_{i}^{I}\bar\omega_{j}^{I}\, ,
\]
with the sum of eigenvalues $\sum\lambda^I=0$.

\item The cross-correlation with dark matter can be non-zero only if statistical isotropy is violated. Assuming that the vector field is fluctuating in some fixed direction $\hat{\omb}$, the cross-correlation takes the form
\begin{align}
\hspace{0.5cm}\corr{\delta_\ms(\kb)\Omega_{i}(\kb')}  &= (2\pi)^{3} W_i P_{\delta\Omega}(k)\delta^{(3)}(\kb +\kb') \,, \\
\mbox{with} \quad W_i &\equiv \hat\omega_i -  \hat{k}_i\hat{k}_j\hat\omega^j  \,. \notag
\end{align}
This form follows from the fact that $\Omega_i$ is a pure vector field i.e.\ divergence free. A non-vanishing  $\corr{\delta_\ms\Omega_{i}}$ always defines a preferred spatial direction $\hat\omega_i$ and therefore violates statistical isotropy.	
		
\end{enumerate}

\begin{figure*}[t]
\includegraphics[width=\columnwidth]{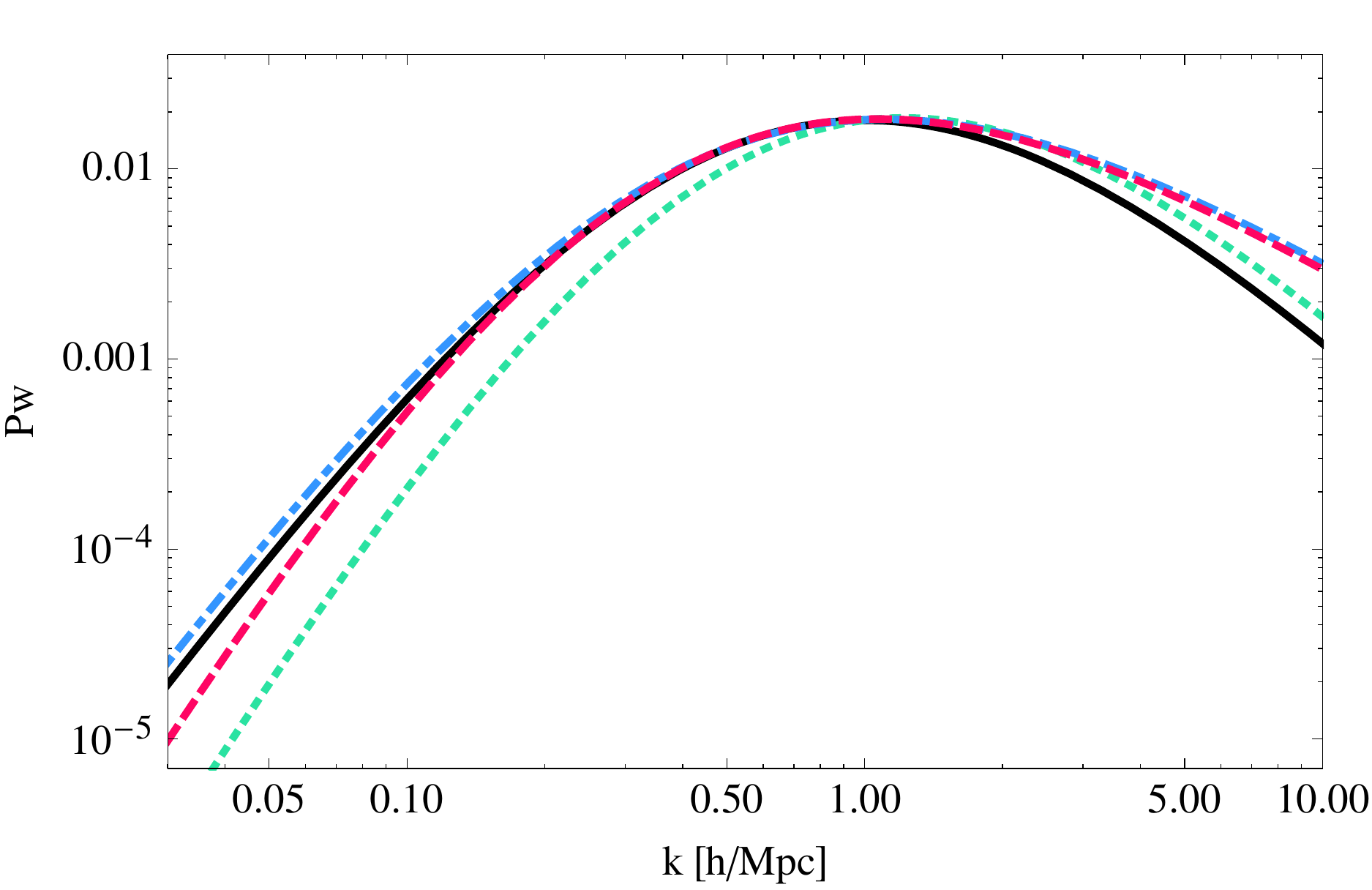}\hspace{0.5cm}\includegraphics[width=\columnwidth]{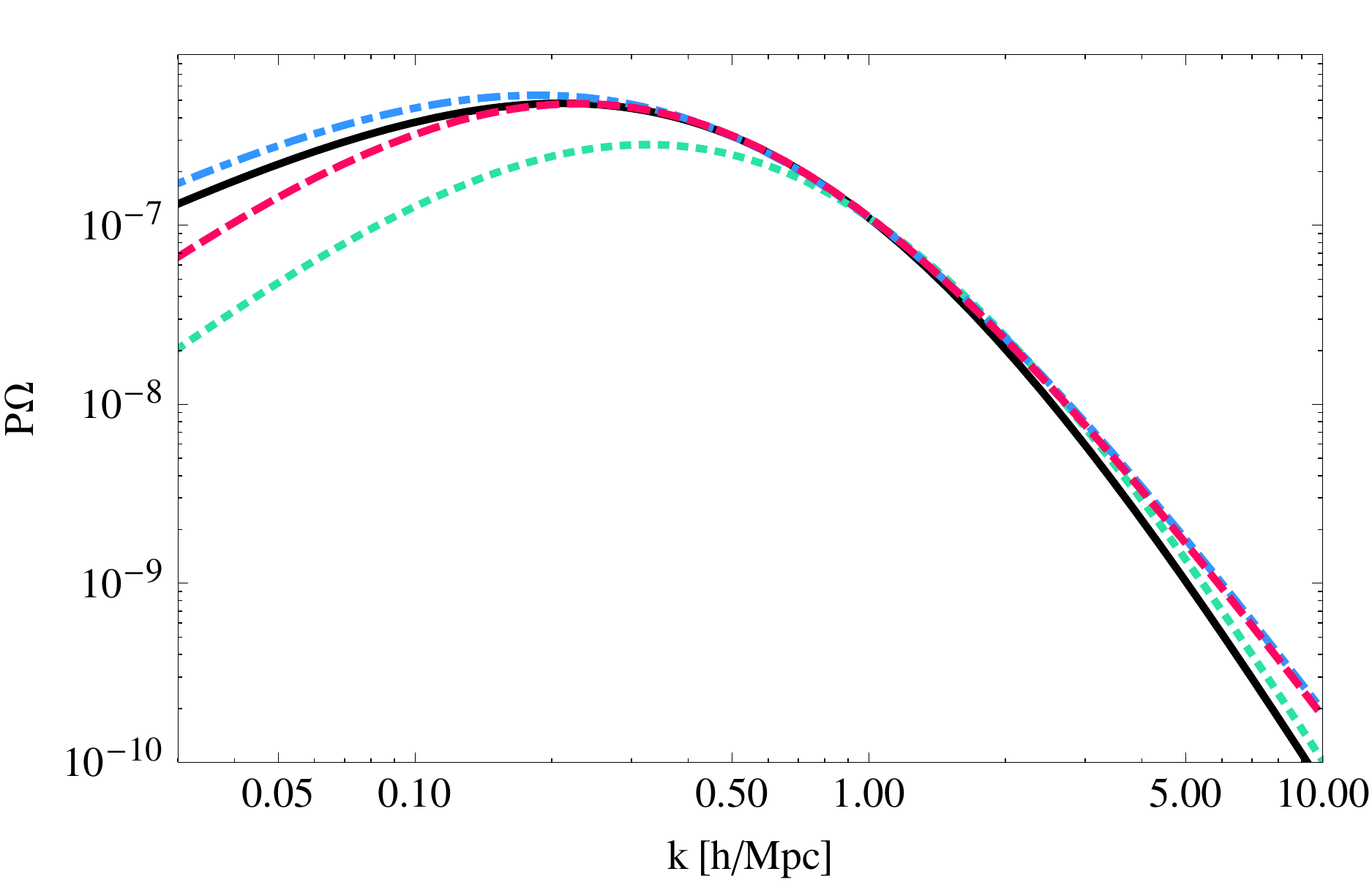}
\caption{\label{f:spectrum}{\it Left panel}: The power spectrum $P_w(k,z=0)$ in units of $({\rm Mpc}/h)^3$ normalised by $(\mathcal{H}_0f_0)^2$, as in Fig. 4 of~\cite{Pueblas:2008uv} (see text for more detail). The black solid line corresponds to the shape defined in Eq.~\eqref{Pom1343}, the green dotted line to Eq.~\eqref{Pom243}, the blue dot-dashed line to Eq.~\eqref{Pom1335} and the red dashed line to Eq.~\eqref{Pom235}. {\it Right panel}: The corresponding power spectrum $P_\Omega(k,z=0)$ in unit $({\rm Mpc}/h)^3$, with the same color coding as in the left panel.}
\end{figure*}

\subsection{Simple models of vector perturbations}
\label{sec:models}

Since sources of vector perturbations are not the main topic of this paper, we will  use simple
parameterisations of two basic scenarios: vector perturbations generated from non-linear structure formation (vorticity), and
vector perturbations generated by topological defects (frame dragging). In these models there are usually no parity violating terms so we shall set $P_A=0$ for this study.

\subsubsection{Vorticity}

Ref.~\cite{Pueblas:2008uv} and, more recently,~\cite{Zhu:2017vtj} studied the generation of vorticity from non-linear structure formation using numerical simulations. Here we assume that the averaged distribution of galaxies follows the same trajectories as the dark matter distribution, so that the galaxy velocity field exhibits the same vorticity as the dark matter velocity field analysed in these numerical simulations. We use the vorticity power spectrum plotted in Fig.\ 4 of~\cite{Pueblas:2008uv} to construct the following fit for $P_\Omega$,
\[
P_\Omega (k,z=0)=A_V \frac{(k/k_*)^{n_\ell}}{\left[ 1+ (k/k_*) \right]^{n_\ell+n_s}}\quad \big({\rm Mpc}/h)^3\, , \label{Pom1343}
\]
where the power at large scales is given by $n_\ell=1.3$, the power at small scales by $n_s=4.3$ and the transition scale by $k_*=0.7\,h/$Mpc. From Fig.\ 4 of~\cite{Pueblas:2008uv} we find that the predicted amplitude for $P_\Omega$ is $A_V=10^{-5}$. In Fig.~\ref{f:spectrum} we plot $P_\Omega(k)$ (right panel, black solid line). In the left panel, we show the quantity calculated in numerical simulations $P_w(k)$, where $\boldsymbol{w}=\boldsymbol{\nabla}\times\boldsymbol{v}$ so that   $P_w(k)$ is related to $P_\Omega$ by a factor $k^2$.%
\footnote{Note that in Fig.~\ref{f:spectrum} and in Fig.\ 4 of~\cite{Pueblas:2008uv} $P_w$ is normalised by $(\mathcal{H}_0f_0)^2$ so that it has the same dimensions as $P_\Omega$. Note also that the convention for the power spectrum used in~\cite{Pueblas:2008uv} differs from our convention by a factor $1/(2\pi)^3$, so that $P_\Omega=(2\pi)^3 P_w(k)/k^2$.} %

According to the numerical simulations of Ref.~\cite{Pueblas:2008uv}, the vorticity power spectrum seems to evolve with $\Hcal(z)^2 f(z)^2 D_1(z)^7$ at large scales. At small scales, the evolution has an additional scale-dependence, leading to a suppression of power at small scales at late times, see Fig.\ 4 of~\cite{Pueblas:2008uv}. In the following we will ignore this small-scale dependence and assume that the power spectrum at redshift $z$ is given by~\footnote{Note that the constraints obtained in this way are conservative, because we underestimate the vorticity power spectrum at small scales for large redshift.}
\[
P_\Omega (k)=P_\Omega (k,z=0)\left(\frac{\Hcal(z)f(z)}{\Hcal_0f(z=0)}\right)^2\left(\frac{D_1(z)}{D_1(z=0)}\right)^7\, . \label{Pevol}
\]

In~\cite{Cusin:2016zvu}, the vorticity generation from large-scale structure was calculated by including velocity dispersion using a perturbative approach. At large scales, this analytical result gives a behaviour slightly different from the numerical one of~\cite{Pueblas:2008uv}, scaling with $n_\ell=2$ instead of $n_\ell=1.3$. This also follows from a simple causality argument: in standard structure formation, vorticity vanishes initially and only 
builds up over time by shell crossing. This is a causal process and we therefore expect the vorticity correlation function to be a function with compact support, which vanishes outside the horizon. Therefore, its Fourier transform, the power spectrum, is analytic. Due to the non-analytic pre-factor $(\delta_{ij} -\hat k_i\hat k_j)$ this requires that for small $k$, 
 $P_\Omega (k) \propto k^n$, where $n$ is an even integer. If there is no additional conservation law which forbids this, we expect $n=2$. The value $n=1.3$ is therefore not possible. Of course a numerical determination of this power law on large scales is difficult and it is not very surprising that N-body simulations find a somewhat  different behaviour.
In addition, the analytical calculation of~\cite{Cusin:2016zvu} finds that the vorticity grows as $D_1(z)$ instead of $\Hcal(z)^2f(z)^2D_1(z)^7$.

Finally, let us note that the numerical simulations of~\cite{Zhu:2017vtj} find a shape for the vorticity power spectrum (see Figure 13) which is broad agreement with~\cite{Pueblas:2008uv}. There are however some differences: first the slope of~\cite{Zhu:2017vtj} at small $k$ is slightly less steep than the one of~\cite{Pueblas:2008uv}. Second, the turnover happens at smaller $k$ in~\cite{Zhu:2017vtj}. And finally, the amplitude of the power spectrum is roughly 50-100 times larger in~\cite{Zhu:2017vtj} than in~\cite{Pueblas:2008uv}, i.e. $A_V\sim 10^{-3}$ instead of $A_V\sim 10^{-5}$.

In the following, we will study how the constraints on vorticity change when we vary the parameters in Eqs.~\eqref{Pom1343} and~\eqref{Pevol}, i.e. $n_\ell, n_s, k_*$ and the evolution with redshift. More precisely, we study three additional fits for $P_\Omega$
\begin{align}
&P_\Omega (k,z=0)=A_V \frac{1.45(k/k_*)^{n_\ell}}{\left[ 1+ (k/k_*) \right]^{n_\ell+n_s}}\quad \big({\rm Mpc}/h)^3\, ,\nonumber\\
&{\rm with}\quad n_\ell=2, n_s=4.3\,\, {\rm and}\,\, k_*=0.7\,h/{\rm Mpc}\, .\label{Pom243}\\
&P_\Omega (k,z=0)=A_V \frac{0.88(k/k_*)^{n_\ell}}{\left[1+ (k/k_*) \right]^{n_\ell+n_s}}\quad \big({\rm Mpc}/h)^3\, ,\nonumber\\
&{\rm with}\quad n_\ell=1.3, n_s=3.5\,\, {\rm and}\,\, k_*=0.5\,h/{\rm Mpc}\, .\label{Pom1335}\\
&P_\Omega (k,z=0)=A_V \frac{1.76(k/k_*)^{n_\ell}}{\left[1+ (k/k_*) \right]^{n_\ell+n_s}}\quad \big({\rm Mpc}/h)^3\, ,\nonumber\\
&{\rm with}\quad n_\ell=2, n_s=3.5\,\, {\rm and}\,\, k_*=0.4\,h/{\rm Mpc}\, .\label{Pom235}
\end{align}
Note that we have adjusted $k_*$ and the amplitude in order to always have $P_w$ which peaks around $k=1\,h/$Mpc with an amplitude of $0.01(\Hcal_0f_0)^2\, ({\rm Mpc}/h)^3$ when $A_V=10^{-5}$. The different curves are plotted in Fig.~\ref{f:spectrum}. In addition to these three additional fits, we also consider model~\eqref{Pom1343} where we vary $k_*$ to $k_*=0.6\,h$/Mpc (corresponding to a peak around 0.85\,$h/$Mpc) and $k_*=0.8\,h$/Mpc (corresponding to a peak around 1.15\,$h/$Mpc), keeping the amplitude $0.01(\Hcal_0f_0)^2\, ({\rm Mpc}/h)^3$ (for $A_V=10^{-5}$) fixed.

\subsubsection{Topological defects}

For topological defects, we use that $\Omega_i \rightarrow \Sigma_i$ as discussed in \ref{sec:basics}, i.e.\ that the galaxy velocities are not vortical, but we use the frame dragging effect  $\Sigma_i \propto \delta T_{i(V)}^0 / k^2$. Based on large numerical field-theory simulations of cosmic strings, Ref.~\cite{Daverio:2015nva} derived fitting functions to the various unequal time correlators of the defect energy-momentum tensor. As discussed in the erratum of~\cite{Daverio:2015nva}, the power spectrum of $T_{0i}$ of a causal source is generically proportional to $k^2$ on large scales. Numerically they find that the power spectrum for $\delta T^0_i$ turns over slightly inside the horizon, $k_* \tau \approx 12$, and then decays as $k^{-1.14}$.  As defects are scaling, it is best to express all quantities except for a dimensionful prefactor in terms of $k \tau \equiv x$, so that finally the dimensionless vector velocity power spectrum for cosmic strings is given by
\[
k^3 P_\Omega (k,\tau) \approx 14 (G\mu)^2 \left( \frac{x/12}{1 +(x/12)^{3.14}} \right) \, . \label{POmdef}
\]
$G\mu$ is a dimensionless number that is linked to the symmetry breaking scale (e.g.\ \cite{Urrestilla:2007sf}). For defects formed in a phase transition at the Grand Unified Theory (GUT) scale, it is of the order of $10^{-6}$. Observational constraints from the Cosmic Microwave Background (CMB) limit it to be smaller than about $10^{-7}$ to $10^{-6}$ depending on the model \cite{Ade:2015xua,Lizarraga:2016onn}.

Of course also in scenarios with topological defects at some points non-linearities lead to shell crossing and the additional vorticity generation that we discussed above. But here we consider the new effect specific to topological defects which generates frame dragging already at the level of linear perturbation theory where we may set $\Omega_i-\Sigma_i=0$.

\section{The Kaiser formula in the presence of vectors\label{sec:rsd}}

The observed galaxy number counts are those in redshift space, with the leading correction arising from the Kaiser term \cite{Kaiser:1987qv}
\[
\Delta(\rb)=\delta_\gs(\rb)-\frac{1}{\mathcal{H}}n^{i}\partial_{i}(n^{j}v_{j}(\rb))\,,
\]
with the galaxy velocity $v_i$ and where we define the line-of-sight direction $\nb$ as 
\[
\nb\equiv\frac{\rb}{r}
\]
i.e.\ the unit vector in the direction of the galaxy lying at $\boldsymbol{r}$, with the observer
located at $\boldsymbol{r}=0$.
Splitting the velocity into the scalar and vector parts, as in Eq.~\eqref{eq:vel}, we have
\begin{equation}
\Delta(\rb)=\delta_\gs(\rb)-\frac{1}{\mathcal{H}}n^{i}n^{j}\big(\partial_{i}\partial_{j}v(\rb)+\partial_{i}\Omega_{j}(\rb)\big) \, .\label{eq:deltaz}
\end{equation}
The effects of vector perturbations in the general relativistic number counts were studied in~\cite{Durrer:2016jzq}, where it was found that the dominant effect are in the RSD. Since in the relativistic $C_\ell(z_1,z_2)$'s the RSD cannot easily be extracted, we study here the impact of the vector field on the two-point correlation function of galaxies. In this study we neglect both the sub-dominant vector relativistic corrections from~\cite{Durrer:2016jzq} and the scalar relativistic corrections derived in~\cite{Yoo:2009au,Bonvin:2011bg,Challinor:2011bk}.%
\footnote{Note that depending on the model responsible for the vector field, the scalar relativistic corrections to the correlation function may be of the same order of magnitude as the dominant vector contributions. If this is the case, the scalar relativistic corrections should be included in the modelling of the two-point correlation function. As we will see in Section~\ref{sec:vorticity}, in the case of vorticity, the vector contribution dominates at small scales, where the scalar relativistic corrections are negligible. Note also that our forecasts do not depend on the importance of the relativistic corrections, since in any case the covariance matrix is strongly dominated by density and RSD.} %
The two-point correlation function is given by
\[
\xi(\rb_1,\rb_2,z_1,z_2)=\langle \Delta(\rb_1,z_1)\Delta(\rb_2,z_2) \rangle\, .
\]
Without redshift space distortion, the correlation function is isotropic and depends therefore only on the galaxy separation
\[
x \equiv |\rb_1 - \rb_2|\, ,
\]
and on the mean distance of the pair from the observer $\bar r$, or equivalently its mean redshift $\bar z$
\[
\bar r =\frac{1}{2}(r_1 + r_2), \quad \bar z=\frac{1}{2}(z_1 + z_2)\, .
\]
Redshift space distortions break the isotropy of the correlation function, which consequently depends also on the orientation of the pair with respect to the line-of-sight. In the flat-sky approximation, $\nb_1=\nb_2=\nb$, neglecting evolution between $z_1$ and $z_2$, the scalar correlation function including redshift space distortion  can be written as

\begin{align}
\xi^\text{scalar}(\bar z, x,\boldsymbol{n}\cdot\hat{\boldsymbol{x}}) =& \left(b^2+\frac{2b}{3}f+\frac{1}{5}f^2\right)C_{0}(x) \label{xiscalar}\\
&-\left(\frac{4b}{3}f+\frac{4}{7}f^{2}\right)C_{2}(x)\mathcal{P}_{2}(\boldsymbol{n}\cdot\hat{\boldsymbol{x}}) + \notag\\
& +\frac{8}{35}f^{2}C_{4}(x)\mathcal{P}_{4}(\boldsymbol{n}\cdot\hat{\boldsymbol{x}}) \,, \notag
\end{align}
where $\mathcal{P}_n$ is the Legendre polynomial of degree $n$, $f$ is the growth rate $f\equiv d\ln D_1/d\ln a$ and 
\begin{equation}
C_{n}(x)=\frac{1}{2\pi^{2}}\int\mathrm{d}k\,k^{2}P_{\delta\delta}(\bar z, k)j_{n}(kx) \,.
\end{equation}
Here $j_n$ is the $n$\textsuperscript{th} spherical Bessel function and $P_{\delta\delta}(\bar z, k)$ is the matter power spectrum at the mean redshift $\bar z$. We have made the standard assumption that the galaxy bias $b$ is deterministic and, like the growth-rate $f$ in $\Lambda$CDM, it is scale independent.

The new vector contribution comprises three terms:
\begin{enumerate}
\item Cross-correlation with the density:\newline
We have
\begin{align}
\hspace{1cm}\xi_{\delta\Omega} = -i&\int\!\frac{\ds^{3}k}{(2\pi)^{3}} \Bigg[\frac{b(z_1)}{\Hcal(z_2)}(\nb_1\cdot\kb)(\nb_1\cdot\wb)\\
&-\frac{b(z_2)}{\Hcal(z_1)}(\nb_2\cdot\kb)(\nb_2\cdot\wb)\Bigg]P_{\delta\Omega}(\bar z,k)e^{i\kb\cdot\xb}\, .\label{eq:Dipole?} \notag
\end{align}
Hence in the flat-sky limit, $\nb_1=\nb_2$, if we neglect evolution, the cross-correlation exactly vanishes, even if $P_{\delta\Omega}\neq0$ and isotropy is violated. We thus neglect this contribution here. Note, however, that the cross-correlation would provide a dipole contribution in the case where we correlate two different populations of galaxies. 
\item Cross-correlation with the scalar velocity:\newline
We have
\begin{align}
\hspace{0.8cm}&\xi_{v\Omega}= i\int\!\frac{\ds^{3}k}{(2\pi)^{3}} (\nb_1\cdot\hat\kb)(\nb_2\cdot\hat{\kb}) P_{\delta\Omega}(k)e^{i\kb\cdot\xb}  \\
&\times \left[\frac{f(z_1)}{\Hcal(z_1)} (\nb_1\!\cdot\!\kb)(\nb_2\!\cdot\! \wb)-\frac{f(z_2)}{\Hcal(z_2)}(\nb_2\!\cdot\!\kb)(\nb_1\!\cdot\! \wb) \right]\,. \notag
\end{align}
Also this contribution  vanishes in the flat-sky limit, if we neglect evolution, even in the presence of anisotropy.
\item Auto-correlation:\newline
We obtain
\begin{align}
\hspace{0.8cm}\xi_{\Omega\Omega} = &\frac{1}{\Hcal(z_1)\Hcal(z_2)}\int\!\frac{\ds^{3}k}{(2\pi)^{3}} k^2(\nb_1\cdot\hat\kb)(\nb_2\cdot\hat\kb)\\
&\times n_1^{i}W_{ij}(\hat\kb)n_2^{j}P_{\Omega}(k)e^{i\boldsymbol{k}\cdot\boldsymbol{x}}\, . \nonumber
\end{align}
The auto-correlation has a complicated tensor structure. In the following we will restrict ourself to the case of statistical isotropy $W_{ij}=\delta_{ij}-\kat_i\kat_j$. In the flat-sky approximation, and neglecting evolution we obtain
\begin{align}
\hspace{0.8cm}\xi_{\Omega\Omega} = &\frac{1}{\Hcal^2}\int\!\frac{\ds^{3}k}{(2\pi)^{3}} k^2(\nb\cdot\hat\kb)^2\big(1+(\nb\cdot\hat\kb)^2\big)\\
&\times P_{\Omega}(k)e^{i\boldsymbol{k}\cdot\boldsymbol{x}}\, . \nonumber
\end{align}
Rewriting the $\nb\cdot\hat\kb$ contributions in terms of Legendre polynomial and integrating over the direction of $\kb$, we obtain for the isotropic contribution
\begin{align}
\hspace{0.8cm}\xi^\text{iso}&(\bar z, x,\boldsymbol{n}\cdot\hat{\boldsymbol{x}}) =\frac{2}{15}\mathcal{P}_{0}(\boldsymbol{n}\cdot\hat{\boldsymbol{x}})C_{0}^{\Omega}(x)\label{xivec}\\
&-\frac{2}{21}\mathcal{P}_{2}(\boldsymbol{n}\cdot\hat{\boldsymbol{x}})C_{2}^{\Omega}(x) 
-\frac{8}{35}\mathcal{P}_{4}(\boldsymbol{n}\cdot\hat{\boldsymbol{x}})C_{4}^{\Omega}(x)\,, \notag
\end{align}
with
\begin{equation}
C_{n}^{\Omega}(x)=\frac{1}{2\pi^{2}}\frac{1}{\Hcal^2}\int\mathrm{d}k\,k^{4}P_{\Omega}(k)j_{n}(kx) \,.\label{COmega}
\end{equation}
The unit vector $\hat{\boldsymbol{x}}$ indicates the direction of the vector connecting the two galaxies.

Notice the extra $k^2$ factor multiplying the power spectrum, which is absorbed in the scalar case when the velocity power spectrum is re-expressed in terms of the density power spectrum.
\end{enumerate}

The vector fluctuations contribute to the same multipoles as the scalar fluctuations while the functional dependence remains limited to that of the galaxy separation $x$ and the orientation of the pair w.r.t.\ to the line-of-sight. On the other hand, the relative contributions to the three multipoles differ and therefore it should be in principle possible to simultaneously measure the amplitude of the vector velocity power spectrum together with the growth rate and the bias. Comparing Eqs.~\eqref{xiscalar} and~\eqref{xivec} we see that when $b\sim 1$ and $f\sim 0.5$ the monopole and quadrupole from vorticity are suppressed by about a factor of 10 with respect to the scalar monopole and quadrupole, whereas the hexadecapole from vorticity and from scalar perturbations have the same pre-factor.

\section{Constraints on Vector Fluctuations\label{sec:fisher}}

We now forecast the constraints on vector fluctuations as expected from future redshift surveys. We study the two cases presented in Section~\ref{sec:models}, namely vector perturbations generated from non-linear structure formation, and vector perturbations generated by topological defects. In both cases we assume that we know the shape of the power spectrum and we forecast the constraints on its amplitude: $A_V$ for the vorticity and $(G \mu)^2$ for topological defects. We consider a $\Lambda$CDM universe and we fix the cosmological parameters to the fiducial values of~\cite{2014MNRAS.441...24A}: $\Omega_m = 0.274, h = 0.7, \Omega_b h^2 = 0.0224, n_s = 0.95$ and $\sigma_8 = 0.8$.

\subsection{Constraints on vorticity} 
\label{sec:vorticity}

We first calculate the constraints on vorticity expected from a survey like the future Dark Energy Spectroscopic Instrument (DESI)~\cite{Aghamousa:2016zmz}. The Bright Galaxy DESI survey will observe 10 million galaxies over 14,000 square degrees at redshift $z\leq 0.3$ with spectroscopic redshift accuracy. We split the sample into three thin redshift bins: $0.05 < z < 0.1$, $0.1 < z < 0.2$ and $0.2 < z < 0.3$ that we assume to be uncorrelated. We assume a mean bias of $b = 1.17$ over the whole sample, similar to the one of the main SDSS sample~\cite{Percival:2006gt}. 

In each redshift bin, we measure the amplitude of the monopole, quadrupole and hexadecapole in bins of separation $x_i$.
The Fisher matrix for the amplitude associated with the multipoles $\ell=0,2,4$ is then given by
\[
\mathcal{F}^\ell_{A_V}(\bar z)=\sum_{ij}\frac{\partial \xi_\ell}{\partial A_V}(\bar z, x_i)\big({\rm cov_\ell^{-1}} \big)(\bar z, x_i,x_j)\frac{\partial \xi_\ell}{\partial A_V}(\bar z, x_j)\, .\label{FAV}
\]       
Here $\xi_\ell$ denotes the amplitude of the monopole, quadrupole and hexadecapole of the correlation function. Since the standard correlation function~\eqref{xiscalar} is independent of the vorticity amplitude $A_V$, and the vector part depends linearly on $A_V$ through the $C_{\ell}^{\Omega}(x_i)$ (see Eqs.~\eqref{COmega} and~\eqref{Pom1343}) the partial derivatives can easily be performed. The integrand in~\eqref{COmega} scales as $k^{4-n_s}$ at large $k$ and oscillates very rapidly. For $n_s=4.3$ it converges very slowly, whereas for $n_s=3.5$ it even diverges. This divergence is, however, not physical: we are using galaxies as our probes of the velocity field, and thus we are insensitive to any modes on scales smaller than the typical size of a galaxy. To account for this effect, we introduce a window function $W(k)=\exp\big[-\left(k/k_c \right)^2 \big]$ in~\eqref{COmega} which removes scales above $k_c$. We choose $k_c=10$\,Mpc$^{-1}$, corresponding to a length of $0.1$\,Mpc i.e.\ the typical size of a galaxy. We have checked that increasing $k_c$ to $100$\,Mpc$^{-1}$ changes the constraints on $A_V$ by less than 2\%.

The matrix ${\rm cov}_\ell$ denotes the covariance matrix of the multipole $\ell$. The covariance contains contributions from Poisson noise and from cosmic variance. Since the scalar correlation function is expected to be much larger than the vector correlation function, we neglect the covariance due to the latter and calculate only the cosmic variance of the scalar part, which strongly dominates the error. We follow the method developed in~\cite{Bonvin:2015kuc,Hall:2016bmm}. The detailed expression for the covariance is given in Appendix~\ref{sec:covariance}.

In Eq.~\eqref{FAV}, the sum runs over all the pixels' separations. We choose a pixel size of 2\,Mpc$/h$, and use pixel separations that are multiples of 2\,Mpc$/h$. Since the signal from vorticity quickly decreases with separation, the constraints strongly depend on the minimum separation that we include in our forecasts. Using  Eq.~\eqref{Pom235} to model the shape of the power spectrum, we find a precision on $A_V$ of $\sigma_{A_V}=6\times 10^{-6}$, combining all three multipoles and using a minimum separation $x_{\rm min}=2$\,Mpc$/h$. This degrades to $\sigma_{A_V}=4.6\times 10^{-4}$ if we increase the minimum separation to $x_{\rm min}=10$\,Mpc$/h$. In both cases we use a maximum separation of 100\,Mpc$/h$. The constraints from individual multipoles are summarised in Table~\ref{tab:DESI}. In both cases the constraints come mainly from the hexadecapole, which is significantly more sensitive than the monopole and quadrupole to the presence of vorticity. At small separations, the vector part of the hexadecapole is indeed 20 to 50 times larger than the vector part of the monopole and 5 to 30 times larger than the vector part of the quadrupole. This is due to the spherical Bessel function $j_4$ in Eq.~\eqref{COmega}, which seems to have a better overlap with the form of the vorticity power spectrum than $j_0$ and $j_2$. Since the covariance of the hexadecapole is of the same order as the covariance of the monopole and of the quadrupole, this results in stronger constraints on vorticity from the hexadecapole.%
\footnote{One would naively expect that since the hexadecapole from scalars is significantly smaller than the monopole and quadrupole from scalars, the covariance of the hexadecapole would also be smaller than the covariance of the monopole and of the quadrupole, resulting in even stronger constraints from the hexadecapole. This is however not the case, since the covariance of the hexadecapole is affected by the modes contributing to the monopole and quadrupole. As a consequence, the covariances of the three multipoles are of the same order of magnitude.} %

We also study the dependence of the constraints on the shape of the power spectrum, using models~\eqref{Pom1343}, \eqref{Pom243} and~\eqref{Pom1335}, instead of \eqref{Pom235}. We find that the constraints depend only mildly on the shape. Choosing $x_{\rm min}=2$\,Mpc$/h$, they change by 14\,$\%$: the best constraints are for models~\eqref{Pom243} to~\eqref{Pom235} and the worst for model~\eqref{Pom1343}. Using $x_{\rm min}=10$\,Mpc$/h$ the constraints change by 37\,$\%$ when varying the shape: the best constraints are in this case for models~\eqref{Pom1335} and~\eqref{Pom235} and the worst for model~\eqref{Pom243}. We then change the position at which the power spectrum peaks, keeping the amplitude fixed. In model~\eqref{Pom1343} we vary $k_*=0.7\,h/$Mpc to $k_*=0.8\,h/$Mpc and  $k_*=0.6\,h/$Mpc. We find that the constraints change by 29\,\% when  $x_{\rm min}=2$\,Mpc$/h$ and by 38\,\% when $x_{\rm min}=10$\,Mpc$/h$. Our forecasts are therefore quite robust with respect to small changes in the shape of the power spectrum. Finally, we vary the evolution of the power spectrum with redshift, using the analytical evolution $D_1(z)$ instead of the numerical evolution of~\eqref{Pevol}, $\Hcal^2(z)f^2(z)D_1^7(z)$. We find that the constraints improve by 40\,\% with the analytical evolution.

The numerical simulations of~\cite{Pueblas:2008uv} find an amplitude for the vorticity $A_V\sim 10^{-5}$, whereas the simulations of~\cite{Zhu:2017vtj} find an amplitude of the order of $A_V\sim 10^{-3}$.  Our results show that in the first case, the presence of vorticity will be measurable only at small separations $x_i < 10$\,Mpc$/h$. On the other hand if the amplitude is a few $10^{-4}$, vorticity will leave an observable impact on scales of 10\,Mpc$/h$ and slightly above. 

\begin{figure*}[th]
\includegraphics[width=9.5cm]{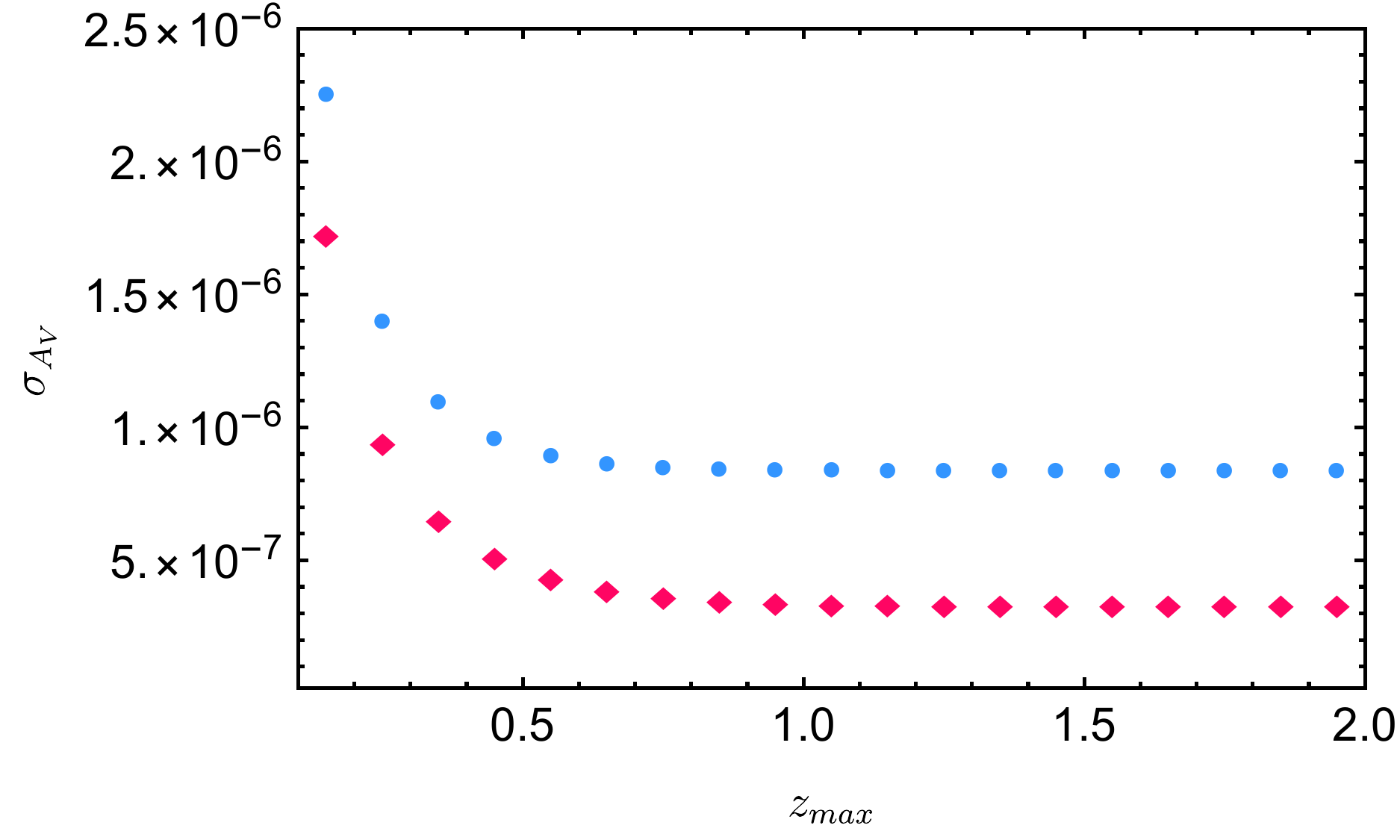}
\caption{\label{f:zmax} Constraints for model~\eqref{Pom235} on the amplitude $A_V$ from the combined monopole, quadrupole and hexadecapole in the SKA, using a mimimum separation $x_{\rm min}=2$\,Mpc$/h$. The constraints are plotted as a function of the maximum redshift bin included in the forecast. The blue dots assume that the vorticity power spectrum evolves as $\Hcal^2(z)f^2(z)D_1^7(z)$, as found in numerical simulations~\cite{Pueblas:2008uv}, whereas the red diamonds assume that the vorticity evolves as $D_1(z)$ as found in the analytical derivation of~\cite{Cusin:2016zvu}.}
\end{figure*}

\begin{figure*}[th]
\includegraphics[width=8.5cm]{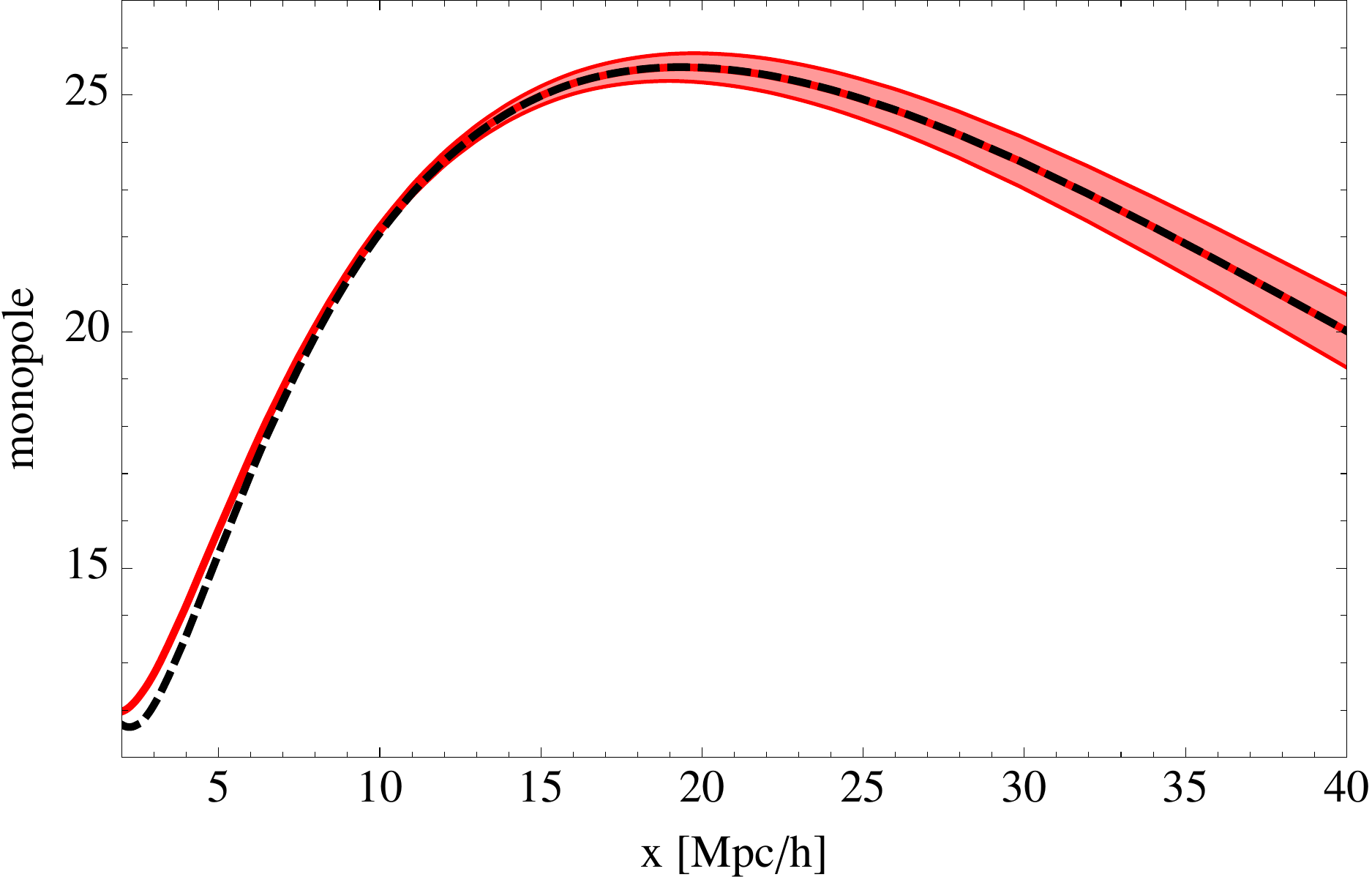}\hspace{0.5cm}\includegraphics[width=8.5cm]{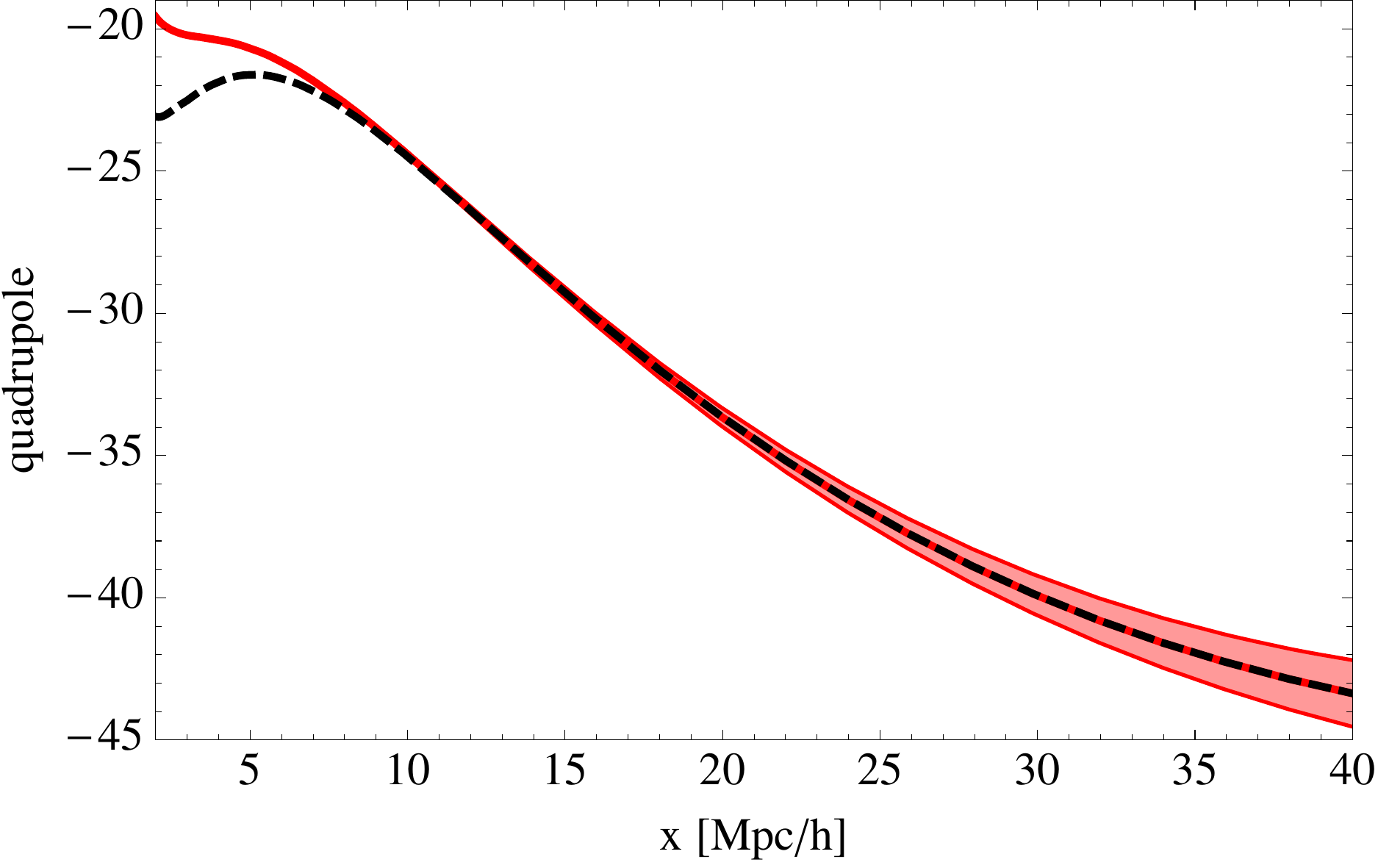}
\caption{\label{f:monoquad} The monopole (left panel) and quadrupole  (right panel) from SKA at $\bar z=0.35$, multiplied by $x^2$ and plotted as a function of separation. The red solid line shows the non-linear scalar multipoles (using halo-fit), with error bars obtained from Eqs.~\eqref{cov0} to~\eqref{cov4}. The black dashed line shows the sum of the non-linear scalar multipoles and the vector contributions with $A_V=5\times 10^{-3}$. Note that the vector contribution is negative for $x\leq 14$\,Mpc/$h$ and positive at larger separations.}
\end{figure*}

\begin{figure*}[th]
\includegraphics[width=8.5cm]{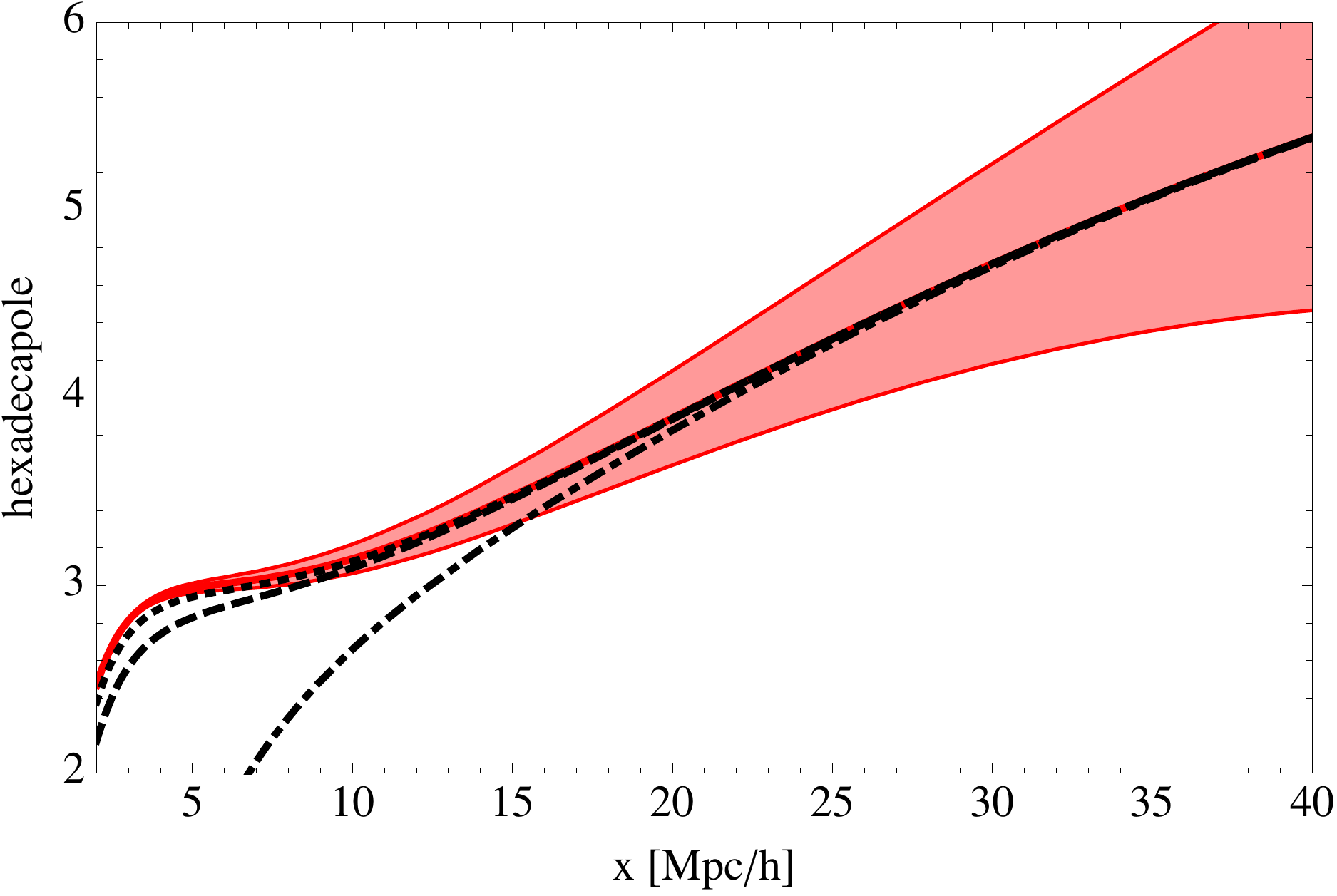}\hspace{0.5cm}\includegraphics[width=8.5cm]{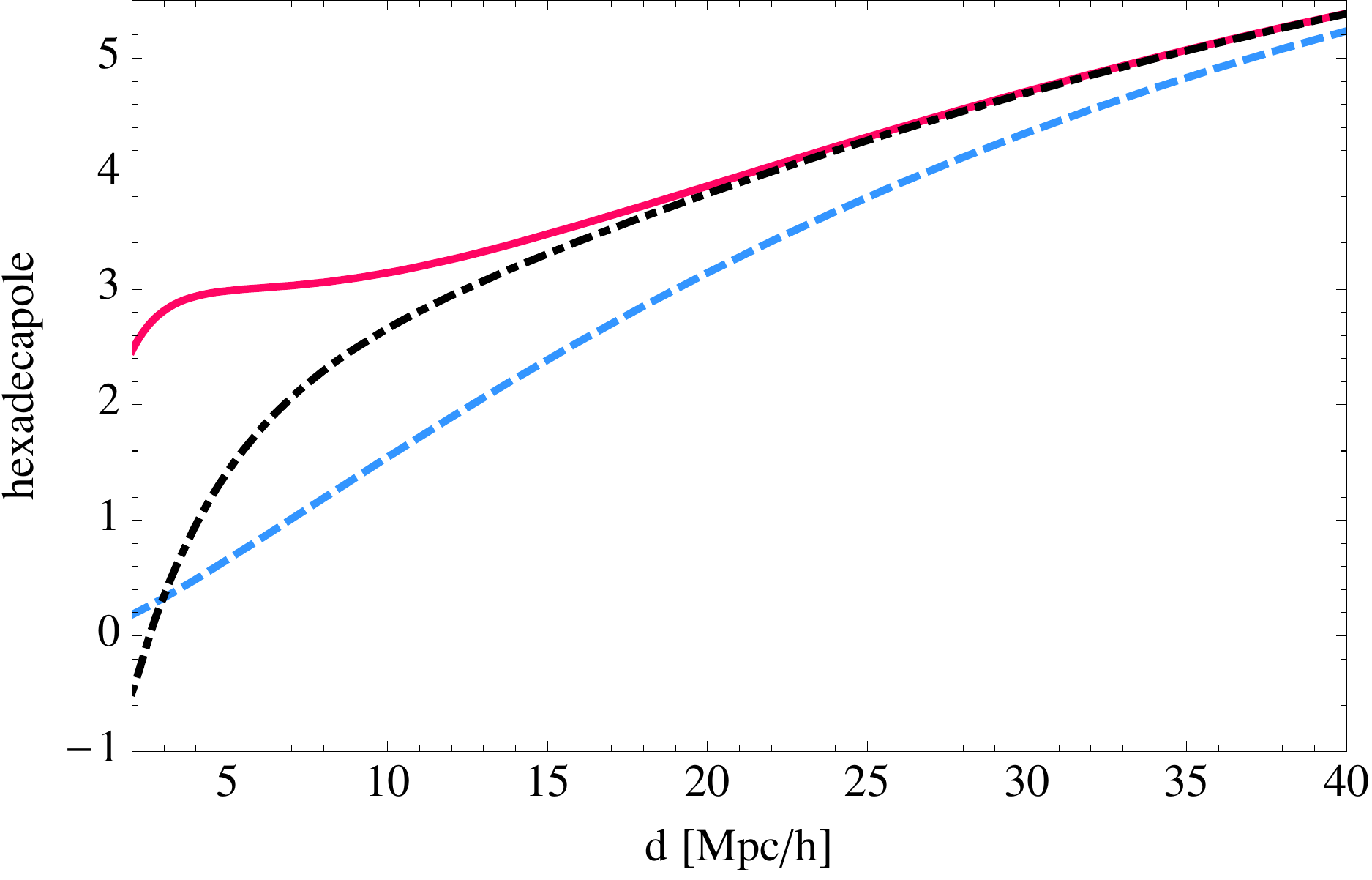}
\caption{\label{f:hexa} The hexadecapole from SKA at $\bar z=0.35$, multiplied by $x^2$ and plotted as a function of separation. {\it Left panel}: the red solid line shows the non-linear scalar contribution to the hexadecapole (using halo-fit), with error bars obtained from Eqs.~\eqref{cov0} to~\eqref{cov4}. The black lines show the sum of the non-linear scalar hexadecapole and the vector contribution with $A_V=3\times 10^{-5}$ (dotted line), $A_V=10^{-4}$ (dashed line) and $A_V=10^{-3}$ (dot-dashed line). {\it Right panel}: the red solid line shows the non-linear scalar contribution to the hexadecapole (using halo-fit), the black dot-dashed line shows the sum of the non-linear scalar hexadecapole and the vector contribution with $A_V=10^{-3}$ and the blue dashed line shows the linear scalar hexadecapole. Note that the vector contribution to the hexadecapole is always negative.}
\end{figure*}

\vspace{1em}

Let us also forecast the constraints obtained from a survey like the SKA. In its second phase of operation the SKA HI (21cm) galaxy survey will detect galaxies spectroscopically from redshift 0.1 to 2 over 30,000 square degrees. We split the redshift range into bins of width $\Delta z=0.1$ and we forecast the constraints on $A_V$ in each of the bins. We use the number density and bias specifications from~\cite{Bull:2015lja} (see Table 3). We then calculate the cumulative constraint from all bins, assuming that they are independent.  Again we choose model~\eqref{Pom235}, since it follows the analytical shape of~\cite{Cusin:2016zvu} at large scales and gives similar constraints as model~\eqref{Pom1343} which fits well the shape found in~\cite{Pueblas:2008uv}. In Table~\ref{tab:SKA} we summarise the constraints on $A_V$ from the individual multipoles as well as the total. We choose different minimum separations: 2\,Mpc$/h$, 10\,Mpc$/h$ and 20\,Mpc$/h$. For all cases the maximum separation is 40\,Mpc$/h$. We have checked that including larger separations does not improve the constraints. 

\begin{table}[t]
\centering
\begin{tabular}{ | c | c  | c | c | c | }
\hline
$x_{\rm min}$ $[{\rm Mpc}/h]$ & mono & quad & hexa & total \\ \hline
2  & $2.5\times 10^{-4}$ & $2.9\times 10^{-5}$ & $6.2\times 10^{-6}$ & $6\times 10^{-6}$ \\ \hline
10 & $6.4\times 10^{-3}$ & $1.3\times 10^{-2}$ & $4.6\times 10^{-4}$ & $4.6\times 10^{-4}$ \\ \hline
\end{tabular}
\caption{Constraints for the model given in Eq.~\eqref{Pom235} with growth $\propto \Hcal(z)^2f(z)^2D_1(z)^7$ on the amplitude $A_V$ from the monopole, quadrupole, hexadecapole, and combined multipoles in DESI. We use separations $x_{\rm min}\leq x_i\leq 100$\,Mpc$/h$ with two values for $x_{\rm min}$.}
\label{tab:DESI}
\end{table}

\begin{table}[t]
\centering
\begin{tabular}{ | c | c  | c | c | c | }
\hline
$x_{\rm min}$ $[{\rm Mpc}/h]$ & mono & quad & hexa & total \\ \hline
2  & $3.7\times 10^{-5}$ & $4.2\times 10^{-6}$ & $8.7\times 10^{-7}$ & $8.5\times 10^{-7}$ \\ \hline
10 & $9.4\times 10^{-4}$ & $2\times 10^{-3}$ & $7.1\times 10^{-5}$ & $7.1\times 10^{-5}$ \\ \hline
20 & $7.2\times 10^{-2}$ & $4.6\times 10^{-2}$ & $1.6\times 10^{-3}$ & $1.6\times 10^{-3}$ \\ \hline
\end{tabular}
\caption{Constraints from model~\eqref{Pom235} with growth $\propto \Hcal(z)^2f(z)^2D_1(z)^7$ on the amplitude $A_V$ from the monopole, quadrupole, hexadecapole, and combined multipoles in the SKA. We use separations $x_{\rm min}\leq x_i\leq 40$\,Mpc$/h$ with three values for $x_{\rm min}$.
\label{tab:SKA}}
\end{table}

\begin{table}[t]
\centering
\begin{tabular}{ | c | c  | c | c | c | }
\hline
$x_{\rm min}$ $[{\rm Mpc}/h]$ & mono & quad & hexa & total \\ \hline
2  & $1.2\times 10^{-5}$ & $1.6\times 10^{-6}$ & $3.5\times 10^{-7}$ & $3.4\times 10^{-7}$ \\ \hline
10 & $3.2\times 10^{-4}$ & $6.6\times 10^{-4}$ & $2.4\times 10^{-5}$ & $2.4\times 10^{-5}$ \\ \hline
20 & $2.2\times 10^{-2}$ & $1.4\times 10^{-2}$ & $5.2\times 10^{-4}$ & $5.2\times 10^{-4}$ \\ \hline
\end{tabular}
\caption{Constraints from model~\eqref{Pom235} combined with the analytical prediction for the growth $\propto D_1(z)$, on the amplitude $A_V$ from the monopole, quadrupole, hexadecapole, and combined multipoles in the SKA. We use separations $x_{\rm min}\leq x_i\leq 40$\,Mpc$/h$ with three values for $x_{\rm min}$.
\label{tab:SKAanalytic}}
\end{table}

From Table~\ref{tab:SKA} we see that if the amplitude of the vorticity power spectrum is of the order of $10^{-5}$ as suggested by the numerical simulations of Ref.~\cite{Pueblas:2008uv},  it should leave an observational impact on both the quadrupole and the hexadecapole at small separations $x_i < 10$\,Mpc$/h$. An amplitude of $10^{-4}$ would leave an impact on scales of the order of 10\,Mpc$/h$ and an amplitude of $10^{-3}$, as suggested by the simulations of Ref.~\cite{Zhu:2017vtj}, on scales as large as 20\,Mpc$/h$. 

In Table~\ref{tab:SKAanalytic}, we repeat the constraints, but this time assuming that vorticity grows linearly with $D_1(z)$, as predicted by the analytical calculation of~\cite{Cusin:2016zvu}. In this case the constraints improve by a factor 2.5 to 3 with respect to those in Table~\ref{tab:SKA}.

Tables~\ref{tab:SKA} and~\ref{tab:SKAanalytic} show the cumulative constraints obtained from combining all redshift bins. Comparing the individual constraints from each bin, we find that using the redshift evolution from numerical simulations $\propto \Hcal^2(z)f^2(z)D_1^7(z)$ the strongest constraint comes from the bin $0.3\leq z\leq 0.4$, whereas using a linear evolution $\propto D_1(z)$ it comes from the bin $0.4\leq z\leq 0.5$. In Figure~\ref{f:zmax} we plot the cumulative constraints as a function of the maximum redshift bin $z_{max}$, for both the numerical and analytical case. We see that in both cases, including bins above $\bar z\simeq1$ does not improve the constraints anymore. This can be understood by noting that the signal decreases with redshift. 
On the other hand, the volume increases with redshift, leading to a decrease of the cosmic variance. Finally the number density decreases with redshift (see Table 3 of~\cite{Bull:2015lja}), leading to an increase of the Poisson noise. Since the constraints come mainly from small separations, the increase of the Poisson noise dominates the error budget at large redshift leading to a decrease in the constraints.

In Figure~\ref{f:monoquad} we plot the monopole and quadrupole at $z=0.35$. We compare the scalar non-linear multipoles, calculated with halo-fit, with the contribution generated by vorticity. At small scales, vorticity leaves an imprint which is larger than the error-bars. As the separation increases, the vorticity contribution quickly decays and the signal is completely dominated by the scalar contribution. 

In Figure~\ref{f:hexa} we show the hexadecapole at $z=0.35$. We compare the hexadecapole from the scalar non-linear contribution (using halo-fit) with the contribution generated by vorticity, for three different values of $A_V$. The impact of vorticity on the hexadecapole is significantly stronger than on the other multipoles. The hexadecapole is therefore ideal to detect the presence of vorticity. In the right panel, we also show the linear scalar contribution. We see that for $A_V=10^{-3}$ (as predicted by~\cite{Zhu:2017vtj}), the contribution from vorticity is similar to the difference between the halo-fit and linear scalar contributions at small scales. At large scales however, the vorticity signal decreases faster than the non-linear scalar contribution. 

Note that in all these plots we have used halo-fit to calculate the non-linear scalar contributions to the multipoles.%
\footnote{More precisely we use the linear continuity equation to relate the velocity to the density and we then calculate the density power spectrum with halo-fit.} %
This does not provide a very accurate description of the scalar velocity in the non-linear regime. More reliable models have been developed to account for the Fingers of God and for the smoothing of the BAO scales, see e.g.~\cite{2013MNRAS.431.2834X}. In this paper we are however not interested in the scalar non-linear signal, but rather in the vector non-linear signal. It is therefore enough for us to use halo-fit to estimate the amplitude of the scalar signal. Note also that our forecasts depend on the form of the scalar contribution only through the covariance matrix, for which halo-fit is sufficiently precise. 

\subsection{Constraints on topological defects}

We now turn to forecast the constraints on topological defects. We use the power spectrum~\eqref{POmdef} to calculate the contribution from the defects to the multipoles and we forecast the constraints expected on the amplitude $(G\mu)^2$. The Fisher matrix for $(G\mu)^2$ is given by
\begin{align}
&\mathcal{F}^\ell_{(G\mu)^2}(\bar z)=\label{Fdefect}\\
&\sum_{ij}\frac{\partial \xi_\ell}{\partial (G\mu)^2}(\bar z, x_i)\big({\rm cov_\ell^{-1}} \big)(\bar z, x_i,x_j)\frac{\partial \xi_\ell}{\partial(G\mu)^2}(\bar z, x_j)\, .\nonumber
\end{align}
Contrary to the vorticity from non-linearities, the $C_n^\Omega$ in Eq.~\eqref{COmega} do not diverge for topological defects, since the power spectrum scales as $k^{-5.14}$ at large $k$. It is therefore not necessary to introduce a window function in this case. We find that the constraints are always much weaker than those obtained from the CMB. We use an optimal survey observing the whole sky between $z=0$ and $z=2$, with a bias evolution similar to that of the SKA. We assume a high number density, so that the Poisson noise can be completely neglected and only cosmic variance contributes to the covariance matrix (first term in Eqs.~\eqref{cov0} to~\eqref{cov4}). We use a pixel size of 2\,Mpc$/h$ and we include separations between 2 and 1000\,Mpc$/h$ in the Fisher matrix~\eqref{Fdefect}. We find that even in this very optimistic settings, the constraints on the amplitude are
\[
\sigma_{(G\mu)^2}=1.3\times 10^{-7}\, ,
\]
which is 6 to 7 orders of magnitude larger than the constraints from the CMB, $(G\mu)^2 < 4 \times 10^{-14}$ \cite{Lizarraga:2016onn,Ade:2015xua}. Redshift space distortions are therefore not competitive to detect the presence of topological defects. This is mainly due to the fact that topological defects leave a very clean imprint on very large scales and at early times that are better probed by the CMB than by large-scale structure.

In this analysis, we have only used linear perturbation theory predictions, based on the result of Eq.\ (\ref{eq:vecgeo}), that the vorticity redshifts away and the velocity field follows the frame being dragged by  metric perturbations (measured in numerical simulations of cosmic strings). In general, we expect non-linear effects like wake-formation behind a cosmic string passing through matter. To properly assess the impact of such non-linear effects we would need to combine numerical string simulations with $N$-body simulations, which goes well beyond the scope of this work. However, it appears unlikely that such effects could increase the vorticity in galaxy velocities by over six orders of magnitude.

\section{Conclusions and Implications}

In this paper, we have shown that the vector contribution to the peculiar velocity of galaxies induces a particular set of corrections to the redshift-space galaxy two-point correlation function. These vector modes arise from either vorticity in the galaxy velocity field or through frame dragging, both affecting the galaxy correlation function. Even when this contribution is isotropic, it is different to the standard scalar one. Even in concordance cosmology, vorticity modes exist since they are produced during structure formation, through shell crossing in the cold dark matter.

We have performed an initial study of the feasibility of detecting this signal and have shown that that next generation surveys such as DESI and SKA will be sensitive enough to detect it. The theoretical uncertainty on the  amplitude of the vorticity power spectrum is still significant, but even at the level of the most pessimistic estimate, a contribution from the vector signal is significant enough to at least become a new systematic for the next generation spectroscopic surveys and should be included in analyses of the correlation function.

We have found that the hexadecapole is most affected by vorticity. At small separations, the vector part of the hexadecapole is indeed 20 to 50 times larger than the vector part of the monopole and 5 to 30 times than the vector part of the quadrupole. Unfortunately, modelling the hexadecapole is well known to be difficult since the effect of the non-linear scalar contribution is relatively larger than for the lower multipoles. Since, for CDM vorticity, the signal is strongest at the smallest separations $x<10$~Mpc$/h$, this is likely to continue being  a challenge. Indeed, deep in the non-linear regime, once perturbation theory has broken down completely, one would naturally expect the signal to be equipartitioned between the fully non-linear scalar and vector modes. A realistic extraction of the vector signal would require a good model for the relationship between the scalar velocity flows and the non-linear matter-density power spectrum, most likely from simulations, with any constraints further degraded by marginalization over nuisance parameters introduced in the models. Nonetheless, the signal is there in the standard model of cosmology, at a level which will affect amplitudes of the correlation function.
	
An exciting possibility is that the signal is actually higher than expected as a result of a non-standard model of dark matter or new degrees of freedom being active in the late universe, or even our inability to properly capture small-scale physics in simulations. For example, in~\cite{Cusin:2016zvu} it was shown that dispersion in the dark matter distribution (i.e.\ if the DM is not completely cold) can create large amplitudes for the vorticity. It is thus in principle possible to use a measurement of the vector modes to put new constraints on the model of dark matter.

Any extended model will come with its own prediction for the shape of the vector power spectrum. We have shown that upcoming surveys are broadly insensitive to  changes to the precise shape and peak position of the vector power spectrum, provided that it is roughly located in the transition between the linear and non-linear regimes. However, the constraints on power spectra peaked at horizon scales, such as for topological defects, are much weaker, mostly as a result of the few galaxy pairs available at such separations and the much larger cosmic variance.

Finally let us mention that vector modes allow one to include the effects of local anisotropy appearing at late times. We have neglected such physics in this work, focussing on isotropic vector fluctuations. The appearance of a preferred direction would have a very specific effect on the correlation function, and would be much better probed using new observables rather than the monopole, quadrupole and hexadecapole of the correlation function into which the full data are currently compressed. We leave the full analysis for future work.

\begin{acknowledgments}
It is a pleasure to thank Mark Hindmarsh for useful discussions.
C.B., R.D. and M.K.~acknowledge
funding by the Swiss National Science Foundation. I.S. is supported by the European
Regional Development Fund and the Czech Ministry of Education, Youth
and Sports (MŠMT) (Project CoGraDS — CZ.02.1.01/0.0/0.0/15\_003/0000437). \end{acknowledgments}

\appendix

\section{Covariance matrix}
\label{sec:covariance}

The covariance of the multipoles $\xi_\ell(x_i)$ contains 3 contributions: a Poisson contribution due to the fact that we observe a finite number of galaxies, a cosmic variance contribution due to the fact that we observe a finite volume and a mixed contribution from Poisson and cosmic variance. Since the scalar contribution to $\xi_\ell(x_i)$ is always larger than the vector contribution at the scales where cosmic variance is important (namely large separations), we can neglect the vector contribution to the cosmic variance and only calculate the scalar contribution. We follow the method developed in~\cite{Bonvin:2015kuc,Hall:2016bmm} to calculate the covariance matrix of the monopole, quadrupole and hexadecapole. We obtain for the monopole,
\begin{align}
&{\rm cov}_0(x_i,x_j)=\label{cov0}\\
&\frac{1}{V}\left(b^4+\frac{4b^3f}{3}+\frac{6b^2f^2}{5}+\frac{4bf^3}{7}+\frac{f^4}{9} \right) D_{\ell=0}\nonumber\\
&+\frac{1}{2\pi \bar N^2 V \ell_p  x_i^2}\delta_{ij}+\frac{1}{\bar N V}\left(b^2+\frac{2bf}{3}+\frac{f^2}{5} \right) G_{\ell=0}\nonumber\, ,
\end{align}
for the quadrupole,
\begin{align}
&{\rm cov}_2(x_i,x_j)=\label{cov2}\\
&\frac{1}{V}\left(\frac{b^4}{5}+\frac{44b^3f}{105}+\frac{18b^2f^2}{35}+\frac{68bf^3}{231}+\frac{83f^4}{1287} \right) D_{\ell=2}\nonumber\\
&+\frac{5}{2\pi \bar N^2 V \ell_p  x_i^2}\delta_{ij}+\frac{1}{\bar N V}\frac{1}{5}\left(b^2+\frac{22bf}{21}+\frac{3f^2}{7} \right) G_{\ell=2}\nonumber\, ,
\end{align}
and for the hexadecapole,
\begin{align}
&{\rm cov}_4(x_i,x_j)=\label{cov4}\\
&\frac{1}{V}\left(\frac{b^4}{9}+\frac{52b^3f}{231}+\frac{1286b^2f^2}{5005}+\frac{436bf^3}{3003}+\frac{79f^4}{2431} \right) D_{\ell=4}\nonumber\\
&+\frac{9}{2\pi \bar N^2 V \ell_p x_i^2}\delta_{ij}+\frac{1}{\bar N V}\frac{1}{3}\left(\frac{b^2}{3}+\frac{26bf}{77}+\frac{643f^2}{5005} \right) G_{\ell=4}\nonumber\, ,
\end{align}
where $V$ denotes the volume of the survey, $\bar N$ is the number density and $\ell_p$ is the size of the cubic pixel in which $\Delta$ is measured. We use $\ell_p=2$\,Mpc$/h$.
The functions $D_\ell$ and $G_\ell$ are defined as follows for $\ell=0,2,4$
\begin{align}
D_\ell(x_i,x_j)=&\frac{(2\ell+1)^2}{\pi^2}\int dk k^2 P_{\delta\delta}^2(\bar z, k)j_\ell(kx_i)j_\ell(kx_j)\, ,\\
G_\ell(x_i,x_j)=&\frac{2(2\ell+1)^2}{\pi^2}\int dk k^2 P_{\delta\delta}(\bar z, k)j_\ell(kx_i)j_\ell(kx_j)\, .
\end{align}
We use halo-fit to calculate the non-linear density power spectrum $P_{\delta\delta}(\bar z, k)$.

\bibliographystyle{utcaps}
\bibliography{VectorRefs}

\end{document}